\documentclass[aps,prd, superscriptaddress,amsmath,amssymb,nofootinbib,nobibnotes,12pt]{revtex4-2}
\usepackage{anyfontsize}
\usepackage{amssymb}
\usepackage{amsmath}
\usepackage{amsfonts}
\usepackage{amsbsy}
\usepackage[utf8]{inputenc} 
\usepackage{newtxtext}
\usepackage{newtxmath}
\usepackage[normalem]{ulem}
\usepackage{tabularray}
\UseTblrLibrary{diagbox}
\usepackage{orcidlink}

\usepackage{tensor}
\usepackage{mathrsfs}
\usepackage{slashed}

\usepackage{bm}

\usepackage[utf8]{inputenc}

\usepackage[dvipsnames]{xcolor}
\usepackage{verbatim,mathtools,needspace,enumitem,etoolbox}%,physics,microtype}
\usepackage{physics}

\usepackage{hyperref}
\hypersetup{
	colorlinks=true,
	urlcolor= blue,
	citecolor=blue,
	linkcolor= blue,
	bookmarks=true,
	bookmarksopen=false,
}

\usepackage{epstopdf}

\usepackage{bm}
\usepackage{dcolumn}

\usepackage{longtable}
\usepackage{enumerate}
\usepackage[normalem]{ulem}

\hypersetup{colorlinks=true,
citecolor=blue,
linkcolor=blue,
urlcolor=blue}
\usepackage{subfigure}
\usepackage{float}
\usepackage{mathrsfs}
\usepackage[dvipsnames]{xcolor}
\usepackage{soul}

\definecolor{darkgreen}{RGB}{1,212,57}

\usepackage{enumitem}

\usepackage{graphicx}% Include figure files
\usepackage{dcolumn}% Align table columns on decimal point
\usepackage{bm}% bold math

\begin{document}

\title{Black Holes Surrounded by Perfect Fluid Dark Matter in Eddington-inspired Born-Infeld Gravity}

\author{A. R. Soares \orcidlink{0000-0003-1871-2068}}
\email{adriano.soares@ifma.edu.br}

\affiliation{Grupo de Investiga\c{c}\~{a}o em Educa\c{c}\~{a}o Matem\'{a}tica, Instituto Federal de Educa\c{c}\~ao Ci\^encia e Tecnologia do Maranh\~ao,  R. Dep. Gast\~ao Vieira, 1000, CEP 65393-000 Buriticupu, MA, Brazil.}

%%%%%%%%%%%%%%%%%%%%%%%%%%%%%%%%%%%%%%%%%%%%%%%%%%%%%%%%%%%%%%%%%%%%%%

\author{C. F. S. Pereira \orcidlink{0000-0001-6913-0223}}
\email{carlosfisica32@gmail.com}
\affiliation{Departamento de F\'isica e Qu\'imica, Universidade Federal do Esp\'irito Santo, Av.Fernando Ferrari, 514, Goiabeiras, Vit\'oria, ES 29060-900, Brazil}

%%%%%%%%%%%%%%%%%%%%%%%%%%%%%%%%

\author{R. L. L. Vit\'oria}
\email{ricardovitoria@professor.uema.br/ricardo-luis91@hotmail.com}
\affiliation{Programa de P\'os-Gradua\c c\~ao em Engenharia Aeroespacial, Universidade Estadual do Maranh\~ao, Cidade Universit\'aria Paulo VI, S\~ao Lu\'is, MA 65055-310, Brazil}
\affiliation{Faculdade de F\'isica, Universidade Federal do Par\'a, Av. Augusto Corr\^ea, Guam\'a, Bel\'em, PA 66075-110, Brazil}


\begin{abstract}
	%%%%%%%%%%%%%%%%%%%%%%%%%%%%%%%%%%%%%%
In this work, we exactly derive the solution for the gravitational field of a black hole in Eddington-inspired Born-Infeld (EiBI) gravity, surrounded by perfect fluid dark matter. We analyze how the event horizon and the black hole dimensions vary as a function of the model parameters, exploring the fundamental properties of this spacetime. Through numerical investigations, we examine the geodesics of massive particles and demonstrate the high sensitivity of stable circular orbits to the system's coupling constants.

\end{abstract}

\maketitle

%\tableofcontents

%%%%%%%%%%%%%%%%%%%%%%%%%%%%

%%%%%%%%%%%%%%%%%%%%%%%%%%%%%%%%%%%%%%
\section{Introduction}\label{sec1}

Einstein's General Relativity (GR) has been the cornerstone of modern physics, with its predictions confirmed by numerous observations, particularly in the regime of weak gravitational fields. However, the theory faces significant challenges in strong-field regimes, such as the presence of curvature singularities at the Big Bang and during gravitational collapse \cite{1457,529}. To overcome such limitations, several modified gravity theories have been proposed. Among them, Eddington-inspired Born-Infeld (EiBI) gravity has garnered interest due to its ability to avoid cosmological singularities and dust collapse singularities \cite{banados2010,deser1998,vollick2004}. 

The structure of the gravitational Lagrangian in this theory is inspired by the non-linear electrodynamics of Born and Infeld \cite{eddington1920mathematical, born1934foundations,reportEiBI}, and is formulated in a metric-affine approach to avoid the higher-derivative equations that typically appear in the standard metric formulation. A distinctive feature of the EiBI theory in this formulation is that it becomes equivalent to General Relativity in a vacuum. Significant deviations from GR arise only in the presence of matter sources, where the energy-momentum tensor is non-vanishing \cite{Olmo:2011uz}. This makes the choice of the matter field a determining factor in exploring the physical properties and novel phenomena predicted by this theory. To date, studies in EiBI have predominantly focused on compact stars, black holes, and wormholes \cite{Nordstrom-olmo, Jana2015, ShaikhNCC, Shaikh2015, soares2019, Soares2020, Olmo:2013gqa, lambaga2018, PhysRevD.90.124087}.

In parallel, the nature of Dark Matter (DM) remains one of the greatest enigmas of contemporary astrophysics. Observations of spiral galaxies indicate that dark matter can contribute approximately $90\%$ of a galaxy's mass \cite{Persic96}. It is widely accepted that astrophysical black holes do not exist in isolation but are immersed in dark matter halos \cite{L1, L6, Boshkayev, ref77,ref74}. Among the various profiles proposed to describe this substance, the Perfect Fluid Dark Matter (PFDM) model stands out for its ability to explain the rotation curves of spiral galaxies in a phenomenologically consistent manner \cite{Kiselev:2003,Kiselev:2003dm, Li:2012, Jha2025}. Considering that EiBI gravity requires the presence of matter to manifest deviations from classical relativity, and given the central role that dark matter plays in the dynamics of compact objects and galaxies, it becomes imperative to investigate how the gravitational modification proposed by EiBI interacts with the dark matter scenario.

In this work, motivated by the need to test alternative theories in strong-field regimes and by the abundance of dark matter in the universe, we propose the derivation of a static and spherically symmetric solution for a black hole in EiBI gravity surrounded by PFDM. This approach aims not only to expand the catalog of exact solutions in modified gravity but also to provide a theoretical basis for future observational tests, such as shadowing and strong gravitational lensing, allowing for the distinction between signatures of EiBI gravity and those imposed by the presence of dark matter.

In Sect.~\ref{sec2}, we solve the EiBI field equations for the PFDM energy-momentum tensor. We then discuss the relationships between the model parameters that enable physically acceptable solutions. In Sect.~\ref{geo}, we derive the geodesics for a massive test particle and investigate the relationship between the innermost stable circular orbit (ISCO) and the model parameters. Finally, we conclude in Sect.~\ref{concl}, highlighting the main findings of this work.
\section{Born-Infield Gravity}\label{sec2}

The  classical action of the Born-Infield gravity is given by \cite{banados2010}
\begin{eqnarray}\label{acao-BI}
	S_{BH}&=&\frac{1}{8\pi \kappa}\int\left[ \sqrt{-|g_{\mu\nu}+\kappa R_{\mu\nu}(\Gamma)|}-\lambda\sqrt{-|g_{\mu\nu}|}\right]d^4x\nonumber\\
	&&+S_m[g_{\mu\nu}, \Phi].
\end{eqnarray}  
Here  $\kappa$ is a constant denominada Eddington parameter. The vertical  bar stands for the matrix determinant. The Ricci tensor $R_{\mu\nu}(\Gamma)$ is  constructed from the connection $\Gamma$  which is independent of the metric tensor $g_{\mu\nu}$  while we do not use the Riemannian formalism. The action $S_{m}[g_{\mu\nu},\Phi]$  describes the dynamics of the metric and the matter fields, represented by $\Phi$. The $\lambda$ is a constant defined as $\lambda=1+\kappa\Lambda $, where $\Lambda$ is the cosmological constant, but in this work we consider the case $\Lambda=0$, hence, $\lambda=1$. In terms of the definitions
\begin{equation}\label{hmn}
	h_{\mu\nu}=g_{\mu\nu}+\kappa R_{\mu\nu},
\end{equation}
the action (\ref{acao-BI}) reads  as
\begin{equation}
	S_{BI}=\frac{1}{8\pi \kappa}\int\left[\sqrt{-h}-\sqrt{-g}\right]d^4x+S_m[g_{\mu\nu},\Phi],
\end{equation}  	
where $h$ and $g$ are, respectively, the determinants of the $h_{\mu\nu}$ and $g_{\mu\nu}$. The variation of the action with respect to $h_{\mu\nu}$ and $g_{\mu\nu}$ leads to field equations 
\begin{equation}\label{EoM1}
	\sqrt{-h}h^{\mu\nu}=\sqrt{-g}\left(g^{\mu\nu}-8\pi\kappa T^{\mu\nu}\right),
\end{equation}
\begin{equation}\label{EoM2}
	\nabla_{\mu}\left(\sqrt{-h}h^{\alpha\beta}\right)=0,
\end{equation}
where the energy-momentum tensor is given by $T^{\mu\nu}=\frac{2}{\sqrt{-g}}\frac{\delta S_m}{\delta g_{\mu\nu}}$. From Eq.~(\ref{EoM2}), we conclude that the connection associated with the auxiliary metric $h_{\mu\nu}$ is $\Gamma_{\mu\nu}^{\alpha}=\frac{1}{2}h^{\alpha\lambda}\left(\partial_\mu h_{\lambda\nu} +\partial_\nu h_{\lambda\mu}-\partial_\lambda h_{\mu\nu}\right)$, i.e.,  the Levi-Civita connection.

As in \cite{PhysRevD.90.124087}, let us take the following \textit{ansatze} for the physical metric ($g_{\mu\nu}$) and the auxiliary metric ($h_{\mu\nu}$):

\begin{equation}
	\begin{aligned} \label{g}
		ds_g^{2} &= -A^{2}(r)f(r)\,dt^{2} + \frac{dr^{2}}{f(r)} \\
		&\quad + r^{2}\left(d\theta^{2} + \sin^{2}\theta\, d\phi^{2}\right) \,
	\end{aligned}
\end{equation}

\begin{equation}
	\begin{aligned}\label{q}
		ds_h^{2} &= -G^{2}(r)F(r)\,dt^{2} - \frac{dr^{2}}{F(r)} \\
		&\quad + H^{2}(r)\left(d\theta^{2} + \sin^{2}\theta\, d\phi^{2}\right) \ .
	\end{aligned}
\end{equation}

In terms of the ansatze, Eqs.~(\ref{g}) and (\ref{q}), we can write Eq.~(\ref{hmn}) as

\begin{align}
	\frac{2}{\kappa F} \left( \frac{A^2 f}{G^2 F} - 1 \right) &= \frac{F''}{F} + \frac{2G''}{G} \nonumber \\
	&+ \frac{3G'F'}{GF} + \frac{2F'H'}{FH} + \frac{4G'H'}{GH} , \label{cmp1} \\[10pt]
	\frac{2}{\kappa F} \left( \frac{F}{f} - 1 \right) &= \frac{F''}{F} + \frac{2G''}{G} \nonumber \\
	&+ \frac{4H''}{H} + \frac{2F'H'}{FH} + \frac{3F'G'}{FG} , \label{cmp2} \\[10pt]
	\frac{1}{\kappa F} \left( \frac{r^2}{H^2} - 1 \right) &= -\frac{1}{H^2 F} + \frac{F'H'}{FH} \nonumber \\
	&+ \frac{H'^2}{H^2} + \frac{H''}{H} + \frac{H'G'}{HG} . \label{cmp3}
\end{align}

In order to write Eq.~(\ref{EoM1}), let us consider the energy-momentum tensor corresponding to anisotropic perfect fluid dark matter \cite{Kiselev:2003, ref77, Jha2025}
\begin{align}\label{materia}
	T_{\mu}^{\nu(\text{DM})} &= \text{diag} [-\rho, p, p, p] \nonumber \\
	&= \text{diag} \left[ \frac{\beta}{8\pi r^3}, \frac{\beta}{8\pi r^3}, -\frac{\beta}{16\pi r^3}, -\frac{\beta}{16\pi r^3} \right] \ ,
\end{align}
with $\beta$ denoting the intensity of the PFDM. The adoption of an energy density for Perfect Fluid Dark Matter (PFDM) proportional to $1/r^3$ introduces, after integrating the field equations, a logarithmic term of the form $\frac{\beta}{r} \ln \frac{r}{|\beta|}$ in the metric potential $f(r)$. In the proposed phenomenology, this term describes the gravitational influence of non-baryonic dark matter, being capable of explaining galactic rotation curves without the need to resort to additional hidden baryonic masses.

With Eqs.~(\ref{g}), (\ref{q}), and (\ref{materia}), we can write Eq.~(\ref{EoM1}) as
\begin{align}
	\frac{A H^2 f}{G r^2 F} &= 1 -\frac{\kappa\beta}{r^2}\label{c}, \\
	\frac{G H^2 F}{A r^2 f} &= 1 -\frac{\kappa\beta}{r^2}\label{c1}, \\
	\frac{G}{A} &=\left(1+\frac{\kappa\beta}{2r^3}\right). \label{c2}
\end{align}

Solving Eqs.~(\ref{c}) through Eq.~(\ref{c2}) for $G$, $H$, and $F$, we obtain

\begin{align}
	H^2 &= r^2-\frac{\kappa\beta}{r} \label{m1}, \\
	G &= A \left(1+\frac{\kappa\beta}{2r^3}\right)  \label{m2},  \\
	F &= \frac{f}{ \left(1+\frac{\kappa\beta}{2r^3}\right)} \label{m3}.
\end{align}
We can combine Eqs.~(\ref{m2}) and (\ref{m3}) to show that
\begin{equation}
	\frac{A}{G}=\frac{F}{f} \label{m4} \ .
\end{equation}
With the aid of Eq.~(\ref{m4}), we can combine Eqs.~(\ref{cmp1}) and (\ref{cmp2}) to show that
\begin{equation}
	\frac{H''}{G} - \frac{G'H'}{G^2} = 0 \rightarrow \left(\frac{H'}{G}\right)' = 0.
\end{equation}
This implies that $G=c_1 H'$, where $c_1$ is a constant. Thus, from Eq.~(\ref{m2}), we can obtain an expression for $A$, namely,
\begin{equation}\label{A}
	A=\frac{c_1H'}{\left(1+\frac{\kappa\beta}{2r^3}\right)} \ .
\end{equation}
Therefore,
\begin{equation}
	A=\frac{c_1}{\sqrt{1-\frac{\beta\kappa}{r^3}}} \ .
\end{equation}
In order to recover the Schwarzschild solution in the limit $\beta\to 0$, it is necessary to set $c_1=1$.
From Eq.~(\ref{cmp3}), we can show that
\begin{align}
	f &= \frac{c_2}{GHH'} \left( 1 + \frac{\kappa \beta}{2r^3} \right) \nonumber \\
	&+ \frac{1}{GHH'} \left( 1 + \frac{\kappa \beta}{2r^3} \right) \int G \left( 1 + \frac{r^2 - H^2}{\kappa} \right) dr ; \label{f01}
\end{align}
where $c_2$ is an integration constant. From the expression for Eq.~(\ref{m1}) and setting $G=c_1 H'$ in Eq.~(\ref{f01}), we obtain

\begin{eqnarray}\label{f02}
	f(r)&=& -\frac{4M\sqrt{r^4-r\kappa\beta}}{2r^3+\beta\kappa}\nonumber\\
	&&+\frac{2\sqrt{r^4-r\kappa\beta}}{2r^3+\beta\kappa}\int\frac{(r+\beta)(2r^3+\beta\kappa)}{2r^{5/2}\sqrt{r^3-\kappa\beta}} \ dr \ .
\end{eqnarray}
To obtain Eq.~(\ref{f02}), we set $c_2=-2M$ so that the solution reduces to the Schwarzschild solution when $\beta\to 0$. Indeed, in this limit, we have $f(r)=1-2M/r$, as expected, since in a vacuum, EiBI reduces to GR when approached via the metric-affine formalism. On the other hand, asymptotically as $r\to \infty$, or when $\beta$ is very small, the function $f(r)$ tends to
\begin{equation}\label{aprx}
	f(r) \approx 1 - \frac{2M}{r} + \beta \left[ \frac{1}{r} \ln \left( \frac{r}{|\beta|} \right) - \frac{3\kappa}{2r^3} + \frac{2M\kappa}{r^4} \right] \ .
\end{equation}

This result indicates that the deviations between EiBI gravity and GR are less perceptible in asymptotic, weak-field regimes. It is worth noting that, since the logarithmic term decays more slowly than the Schwarzschild ($1/r$) and EiBI ($1/r^3$) terms, it dominates the long-range behavior of gravity in this model.

After integrating Eq.~(\ref{f02}), we obtain
\begin{align}
	f(r) &= \frac{2 (3r + \beta) \sqrt{(r^3 - \beta \kappa)(r^4 - r \beta \kappa)}}{3 r^{3/2} (2 r^3 + \beta \kappa)} \nonumber \\
	&+ \frac{4 \beta \sqrt{r^4 - r \beta \kappa}}{3 (2 r^3 + \beta \kappa)} \operatorname{ArcTanh} \left[ \frac{r^{3/2}}{\sqrt{r^3 - \beta \kappa}} \right] \nonumber \\
	&- \frac{4 M \sqrt{r^4 - r \beta \kappa}}{2 r^3 + \beta \kappa} . \label{f1}
\end{align}

As we can see, given the dependence of $f(r)$ on the function $\operatorname{ArcTanh}$, we can easily argue that the solution can only be real for $\beta\kappa<0$. This implies that $\beta$ and $\kappa$ must have opposite signs. The solution given by Eq.~(\ref{f1}) exhibits a divergence arising from the $\mathrm{ArcTanh}$ term when its argument approaches unity, which occurs at $r_{div}=(|\beta\kappa|/2)^{1/3}$ for $\beta\kappa<0$. However, for most choices of the parameters this radius lies inside the event horizon. In this case, the divergence is hidden behind the horizon and therefore does not affect the exterior spacetime accessible to distant observers. Moreover, since the present solution arises within a modified gravity framework that can be regarded as an effective description, it is expected that additional corrections may become relevant at sufficiently high curvatures. Thus, the appearance of such an internal divergence may simply signal the breakdown of the effective description in the deep interior region, without compromising the physical validity of the exterior solution. Next, we will consider the behavior of $f(r)$ for cases where we can have a black hole solution, that is, with $\kappa\beta<0$. Beforehand, let us state that, according to \cite{Pani2011}, the existence of neutron stars implies constraints on the Eddington parameter, $\kappa$, namely,
\begin{equation}
	\left| \frac{\kappa}{M^2} \right| \lesssim 6.87 \times 10^3 \times \left( \frac{M_{\odot}}{M} \right)^2,
\end{equation}
where $M_{\odot}$ corresponds to the solar mass and $M$ is the mass of the PFDM-EiBI black hole.

\begin{figure}[h]
	\centering
	\includegraphics[scale=0.59]{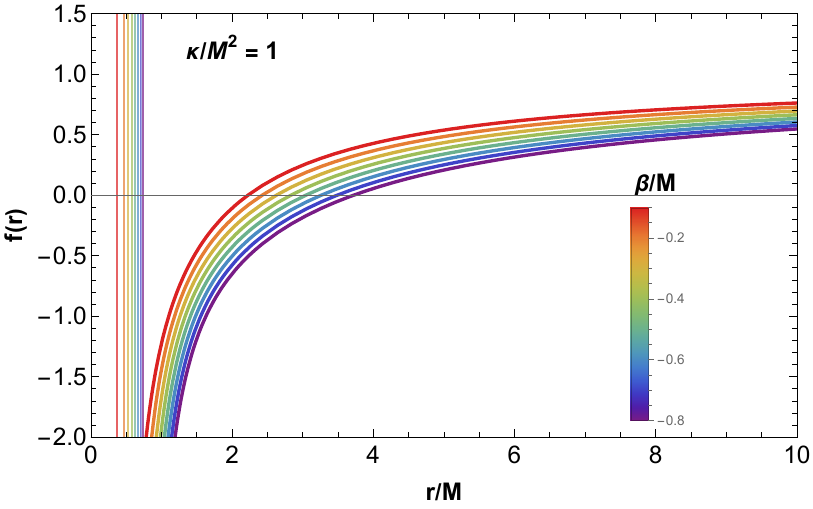}
	\includegraphics[scale=0.59]{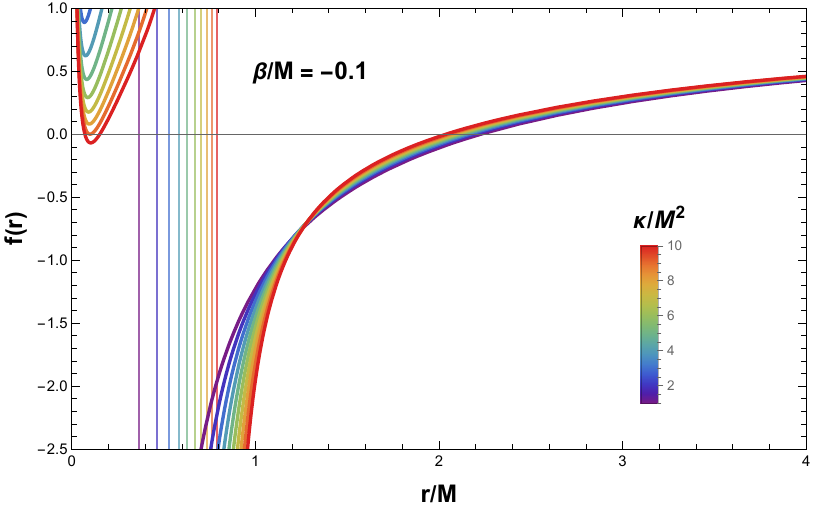}
	\caption{Behavior of the function $f(r)$ for $\kappa/M^2 > 0$.}
	\label{kp}
\end{figure}

 In Fig. \ref{kp}, we show the behavior of the function $f(r)$ in terms of $r/M$ for $\kappa>0$. In the left panel, we set $\kappa/M^2=1$ and plot the function $f(r)$ for various values of $\beta/M<0$. In the right panel, we choose several values for $\kappa/M^2$ while keeping $\beta/M=-~0.1$ fixed. In the panels, the positions of the singularities are marked by vertical lines. Note that the singularity is always covered by the event horizon, at least for the chosen parameter sets, where the event horizon is determined by $f(r)=0$.

\begin{figure}[h]
	\centering
	\includegraphics[scale=0.59]{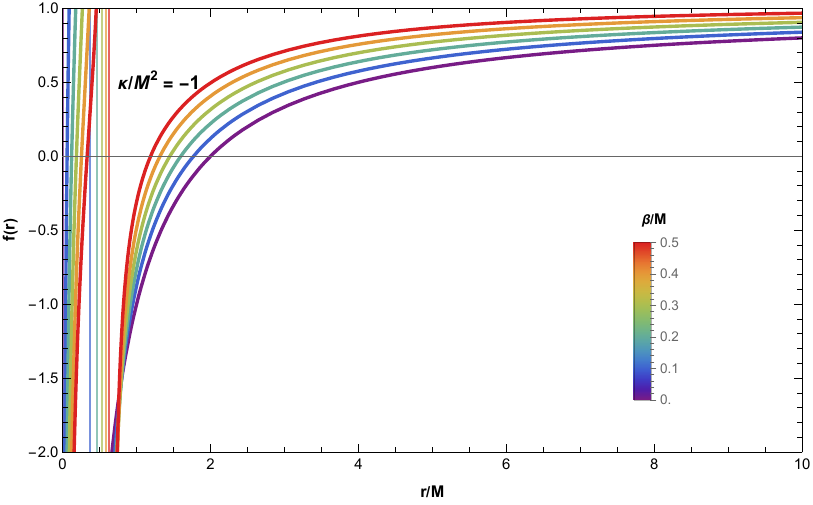}
	\includegraphics[scale=0.59]{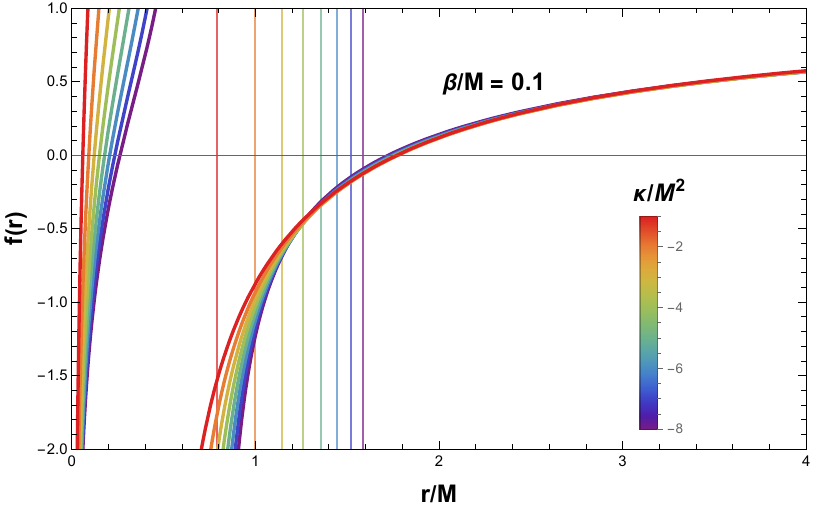}
	\caption{Behavior of the function $f(r)$ for $\kappa/M^2 < 0$.}
	\label{kn}
\end{figure}

Analogously, in Fig.~\ref{kn}, we show the behavior of the function $f(r)$ for $\kappa < 0$. In the left panel, we fix $\kappa/M^2 = -1$ and plot the function $f(r)$ for various values of $\beta/M > 0$. In the right panel, we choose several values for $\kappa/M^2$ and keep $\beta/M = 0.1$ fixed. In general, it is worth noting that when $\kappa > 0$, the radius of the event horizon in EiBI theory is larger than in the Schwarzschild case, whereas for $\kappa < 0$, it becomes smaller.

% Observe que no painel do lado direito, a singularidade aproxima-se cada vez mais do horizonte de eventos à medida que $\kappa/M^2$ torna-se mais negativo. This indicates that there is a parameter regime that does not allow for the existence of black holes

\section{GEODESICS}\label{geo}
Let us consider the geodesic equations of the PFDM-EiBI spacetime. We start by considering the Lagrangian
\begin{equation}
	\mathcal{L} = g_{\mu\nu}\dot{x}^\mu\dot{x}^\nu, \tag{27}
\end{equation}
with $\mathcal{L} = -1, 0, 1$ corresponding to timelike, null, and spacelike geodesics, respectively. The dot denotes the derivative with respect to the affine parameter. Considering the equatorial plane $\theta = \frac{\pi}{2}$, the physical metric Eq.~(\ref{g}) provides, without loss of generality,
\begin{equation}\label{lag}
	\mathcal{L} = -A^2 f \dot{t}^2 + \frac{1}{f}\dot{r}^2 + r^2\dot{\phi}^2. \tag{28}
\end{equation}

The Euler-Lagrange equation for the coordinates $t$ and $\phi$ provides the following conserved quantities:
\begin{equation}
	E = A^2 f \dot{t} \quad \text{and} \quad L = r^2 \dot{\phi}. \tag{29}
\end{equation}
In terms of these new quantities, Eq.~(\ref{lag}) can be written as
\begin{equation}
	\dot{r}^2=\frac{E^2}{A^2}+f\left(\mathcal{L}-\frac{L^2}{r^2}\right) \ .
\end{equation}
For massive particles ($\mathcal{L}=-1$), we obtain
\begin{equation}\label{eqr}
	\dot{r}^2=\frac{E^2}{A^2}-f\left(1+\frac{L^2}{r^2}\right) \ .
\end{equation}

Note that this corresponds to the equation of motion of a classical particle of unit mass, total energy $\mathcal{E} = E^2/A^2$, subject to the one-dimensional effective potential
\begin{equation}\label{pot_ef}
	V_{\text{eff}}= f\left(1+\frac{L^2}{r^2}\right) \ .
\end{equation}

\begin{figure}[h]
	\centering
	\includegraphics[scale=0.59]{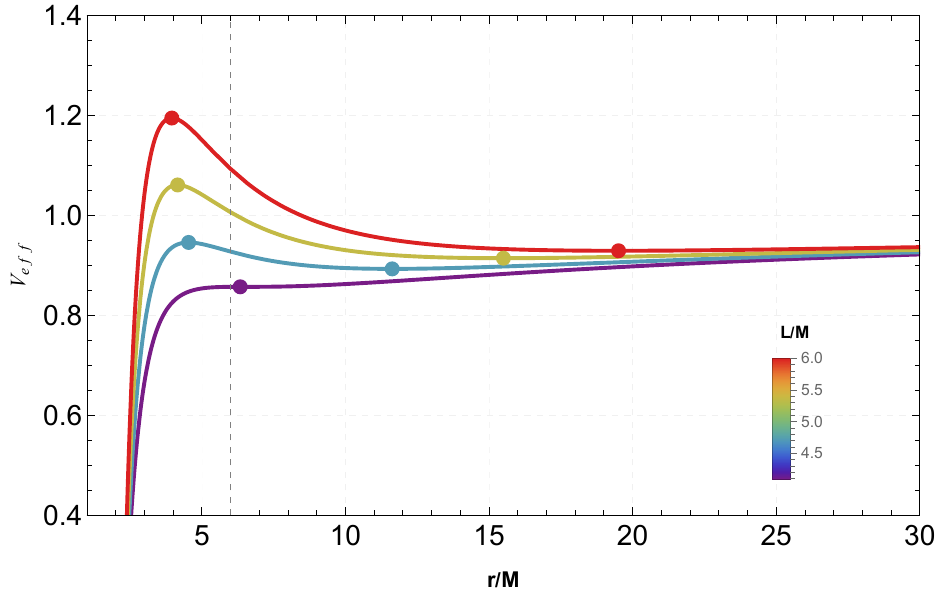}
	\caption{Effective potential, $V_{\text{eff}}$, for $\beta/M = -0.2$ and $\kappa/M^2 = 5$.}
	\label{Vefe}
\end{figure}

In this sense, we can write Eq.~(\ref{eqr}) in the form

\begin{equation}
	\dot{r}^2 = \mathcal{E} - V_{\text{eff}} \ .
\end{equation}
As is well known, radial motion will occur whenever $\mathcal{E} - V_{\text{eff}} > 0$, and turning points arise when $\mathcal{E} = V_{\text{eff}}$. Circular orbits of constant radius can occur in regions where the potential is flat, i.e., $dV_{\text{eff}}/dr = 0$. In this case, we can distinguish between two situations. When $d^2V_{\text{eff}}/dr^2 > 0$, the orbits will be stable, which we call stable circular orbits (SCO); that is, after being slightly displaced from their orbit, the particles quickly return. When $d^2V_{\text{eff}}/dr^2 < 0$, the orbits will be unstable (UCO), and the particles will not return to their orbits after undergoing small perturbations.In Fig.~(\ref{Vefe}), we illustrate the behavior of the effective potential for various values of $L/M$ while keeping $\beta/M=-0.2$ and $\kappa/M^2=5$. Furthermore, we mark for each $L/M$ the position of the SCO, which has a larger radius, and the UCO. As we can observe, just as in GR, as $L/M$ decreases, the SCO moves to the left and the maximum of the potential, which corresponds to the position of the UCO, decreases. Additionally, the SCO and UCO approach each other. Below a certain minimum value of $L/M$, circular orbits cease to exist; the minimum value of $L/M$ for which this occurs determines the innermost stable circular orbit (ISCO), described in Fig.~(\ref{Vefe}) by the purple curve. For the adopted parameters, $L_{ISCO}/M\approx4.1$ and the ISCO radius is $r_{ISCO}/M\approx 6.5$. It is worth noting that in the case of a Schwarzschild BH, $L_{ISCO}/M=2\sqrt{3}\approx3.5$ and $r_{ISCO}/M= 6$; we plot a dotted line indicating the $r_{ISCO}/M$ position corresponding to the Schwarzschild BH. As we can see, it is smaller than in the case of EiBI gravity for the adopted parameters. In any case, it is easy to see that depending on the theory, we can have different values for the radii of the stable circular orbits.

\begin{figure}[h]
	\centering
	\includegraphics[scale=0.59]{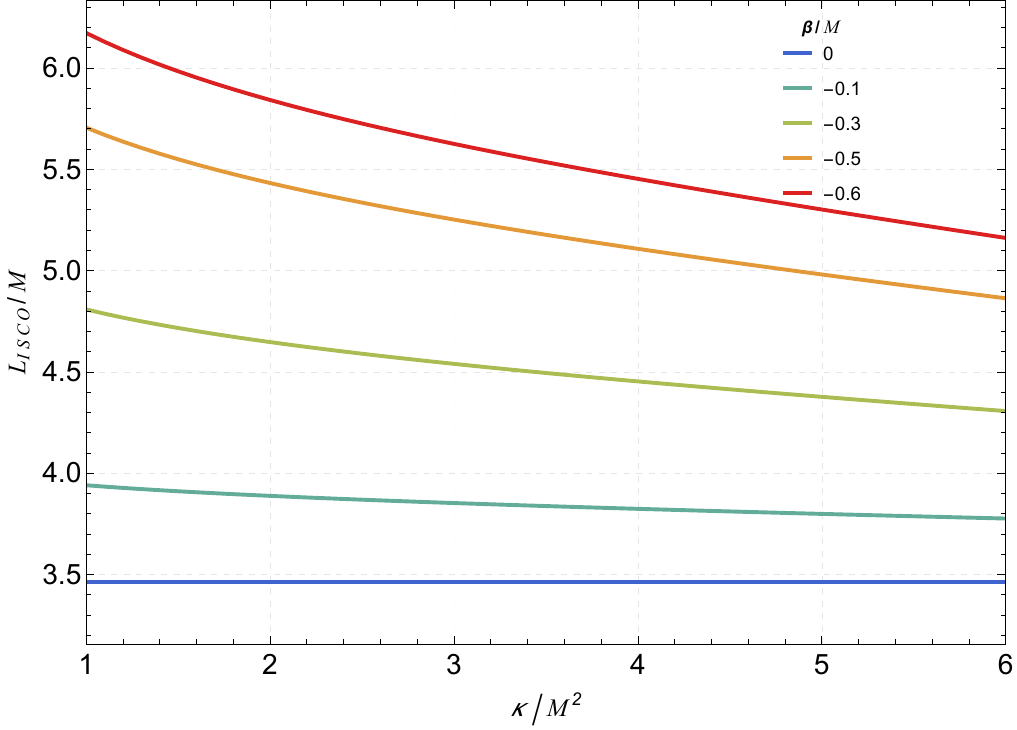}
	\caption{ $L_{ISCO}/M$ for positive values of $\kappa/M^2$. }
	\label{L}
\end{figure}

To clearly illustrate how the metric-affine geometry alters the black hole signature, let us show how $L_{ISCO}/M$ depends on the model parameters. In Fig.~\ref{L}, we present the plot of $L_{ISCO}/M$ as a function of $\kappa/M^2 > 0$, considering some fixed values of $\beta/M < 0$. As can be observed, $L_{ISCO}/M$ increases as the magnitude of $\beta/M$ increases. This means that for a particle to orbit closer to the black hole, it requires more angular momentum. Conversely, it is observed that $L_{ISCO}/M$ presents an inverse correlation with the parameter $\kappa/M^2$. Physically, this behavior suggests that high values of $\kappa$ play a stabilizing role in the system, allowing the particle to maintain a stable circular orbit even under conditions of reduced angular momentum. In the particular case where $\beta/M=0$, we recover the Schwarzschild limit, for which $\frac{L_{ISCO}}{M} = 2\sqrt{3} \approx 3.5$.

\begin{figure}[h!]
	\centering
	\includegraphics[scale=0.59]{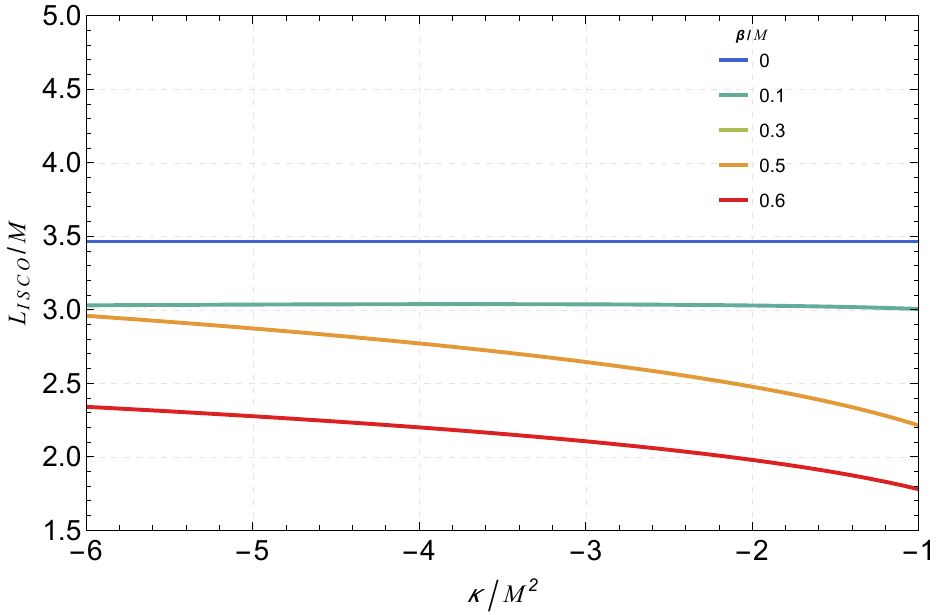}
	\caption{$L_{ISCO}/M$ para valores negativos de $\kappa/M^2$}
	\label{L2}
\end{figure}

In Fig.~\ref{L2}, we present the plot of $L_{ISCO}/M$ as a function of $\kappa/M^2 < 0$, considering some fixed values of $\beta/M > 0$. We observe that as $\beta/M$ increases, $L_{ISCO}/M$ decreases; in this sense, we can state that the PFDM around the black hole tends to stabilize circular geodesics, allowing the particle to maintain its equilibrium with less support from the centrifugal barrier. On the other hand, an increase in the magnitude of $\kappa/M^2$ produces the opposite effect, requiring the particle to have higher angular momentum to counterbalance the modified gravitational attraction and avoid capture by the event horizon.

\section{Conclusion}\label{concl}

In this work, we have derived an exact solution for the gravitational field of a static, spherically symmetric black hole within the framework of Eddington-inspired Born-Infeld (EiBI) gravity, immersed in a perfect fluid dark matter (PFDM) atmosphere. The obtained solution consistently recovers the Schwarzschild vacuum limit ($\beta \to 0$), validating the equivalence between EiBI theory and General Relativity in the absence of matter sources.Our results demonstrate that the existence of physically acceptable solutions imposes a fundamental constraint on the signs of the model parameters, requiring $\kappa$ and $\beta$ to have opposite signs ($\kappa\beta < 0$). We verified that the geometry of the event horizon is significantly affected by the metric-affine coupling: for $\kappa > 0$, the horizon radius is larger than the Schwarzschild radius, while for $\kappa < 0$, the black hole becomes effectively more compact.The investigation of geodesics for massive particles revealed that the orbital dynamics are profoundly shaped by the interaction between modified gravity and dark matter. The analysis of the innermost stable circular orbit (ISCO) indicated that, in the $\beta > 0$ regime, the dark matter fluid acts as a stabilizing agent, reducing the critical angular momentum ($L_{ISCO}/M$) required to maintain orbital equilibrium. Regarding the influence of EiBI gravity, the Eddington parameter $\kappa$ exerts the opposite effect in strong-field regimes; an increase in its magnitude requires greater centrifugal barrier support to prevent the particle from being captured by the event horizon.In summary, this study provides a solid theoretical basis for distinguishing the signatures of modified gravity from those imposed by dark matter environments. The observed deviations in stable circular orbits suggest that future astrophysical tests, such as black hole shadowing and strong gravitational lensing, could be effective tools for constraining the $\kappa$ and $\beta$ parameters and testing the validity of EiBI gravity in strong-field regimes.

\acknowledgments{The author R. L. L. Vit\'oria would like to thank UEMA (Universidade Estadual do Maranh\~ao) and to CNPq (Conselho Nacional de Desenvolvimento Cient\'ifico e Tecnol\'ogico-Brazil). R. L. L. Vit\'oria was supported by the EDITAL N.º 102/2025-PPG/CPG/UEMA and by CNPq Project No. 150420/2025-0.}

%\begin{acknowledgments}
	
%\end{acknowledgments}

%\nocite{*}


\begin{thebibliography}{9}
	
	 \bibitem{1457}	Gravitational collapse and space-time singularities, Phys. Rev. Lett. {\bf 14}, 57 (1965).
	%
	
	\bibitem{529}	S. W. Hawking, and R. Penrose, The singularities of gravitational collapse and cosmology, Proc. R Soc. Lond. Ser A. 314,	529-548 (1970).
	
	\bibitem{deser1998}
	S.~Deser and G.~W.~Gibbons, \textit{Classical Quantum Gravity} \textbf{15}, L35 (1998).
	
	\bibitem{vollick2004}
	D.~N.~Vollick, \textit{Phys. Rev. D} \textbf{69}, 064030 (2004).
	
	\bibitem{banados2010}
	M.~Ba\~{n}ados and P.~G.~Ferreira, \textit{Phys. Rev. Lett.} \textbf{105}, 011101 (2010).
	
	 \bibitem{eddington1920mathematical} A. S. Eddington, The Mathematical Theory of Relativity, (Cambridge University Press, Cambridge, England, 1924).
	
	 \bibitem{born1934foundations} M. Born and L. Infeld, Proc. R. Soc. A 144, 425 (1934).  % Foundations of the new field theory
	
	 \bibitem {reportEiBI}J. B. Jimenez, L. Heisenberg, G. J. Olmo, and D. Rubiera-Garcia, Phys. Rep. {\bf727}, 1 (2018).  %Born-Infeld inspired modifications of gravity
	 
	  \bibitem{Olmo:2011uz} 
	 G.~J.~Olmo,
	 %``Palatini Approach to Modified Gravity: f(R) Theories and Beyond,''
	 Int.\ J.\ Mod.\ Phys.\ D {\bf 20}, 413 (2011)
	 %doi:10.1142/S0218271811018925
	 [arXiv:1101.3864 [gr-qc]].
	 
	 %%%%%%%%%%%%%%%%%%%%%%%%%%%%%%%%%%%%%%%%%
	 
	  \bibitem{Nordstrom-olmo} G.J. Olmo and D. Rubiera-Garcia, Phys. Rev. D {\bf86}  044014 (2012). % Reissner-NordstroÂ¨m black holes in extended Palatini theories
	  
	   \bibitem{Jana2015} S. Jana and S. Kar, Phys. Rev. D {\bf92}, 084004 (2015). % Born-Infeld gravity coupled to Born-Infeld electrodynamics
	 
	   \bibitem{ShaikhNCC} R. Shaikh, Phys. Rev. D {\bf98}, 064033 (2018). % Wormholes with nonexotic matter in Born-Infeld gravity
	
	\bibitem{Shaikh2015} R. Shaikh, Phys. Rev. D {\bf92}, 024015 (2015). % Lorentzian wormholes in Eddington-inspired Born-Infeld gravity
	
	 \bibitem{soares2019} J. R. Nascimento, G. J. Olmo, A. Yu. Petrov, P. J. Porfirio, A. R. Soares,  Phys.\ Rev.\ D {\bf 99}, 064053 (2019). %Global monopole in Palatini f(R) gravity.
	
	\bibitem{Soares2020} J. R. Nascimento, G. J. Olmo, P. J. Porfírio,
	A. Y. Petrov and A. R. Soares, Phys. Rev. D {\bf101}, 064043 (2020). %Nonlinear ? -models in the Eddington-inspired Born-Infeld gravity
	
		\bibitem{Olmo:2013gqa}
	G.~J.~Olmo, D.~Rubiera-Garcia and H.~Sanchis-Alepuz,
	%``Geonic black holes and remnants in Eddington-inspired Born-Infeld gravity,''
	Eur. Phys. J. C \textbf{74}, 2804 (2014).
	
		 
	\bibitem{lambaga2018} R. D. Lambaga and H. S. Ramadhan, Eur. Phys. J. C {\bf 78}, 436 (2018)% Gravitational field of global monopole within the Eddington-inspired Born-Infeld theory of gravity
	
	\bibitem{PhysRevD.90.124087}  H. Sotani and U. Miyamoto, Phys. Rev. D {\bf90}, 124087 (2014)% Properties of an electrically charged black hole in Eddington-inspired Born-Infeld gravity

%%%%%%%%%%%%%%%%%%%%%% Matéria escura %%%%%%%%%%%%%%%%%

\bibitem{Persic96} 
M. Persic et al., Mon. Not. R. Astron. Soc. \textbf{281}, 27 (1996).
%\textit{The Universal Rotation Curve of Spiral Galaxies: I. the Dark Matter Connection}	 

\bibitem{L1} 
K. Akiyama et al., Astrophys. J. \textbf{875}, L1 (2019). % \textit{First M87 Event Horizon Telescope Results. I. The Shadow of the Supermassive Black Hole}

\bibitem{L6} 
K. Akiyama et al., Astrophys. J. \textbf{875}, L6 (2019). %\textit{First M87 Event Horizon Telescope Results. VI. The Shadow and Mass of the Central Black Hole}


\bibitem{Boshkayev} 
K. Boshkayev and D. Malafarina, Mon. Not. R. Astron. Soc. \textbf{484}, 3325 (2019). % \textit{A model for a dark matter core at the galactic center}

\bibitem{ref77} 
Tian-Chi Ma et al., Modern Phys. Lett. A, 2150112 (2021). %\textit{Shadow cast by a rotating and nonlinear magnetic-charged black hole in perfect fluid dark matter}

\bibitem{ref74} 
K. Saurabh and K. Jusufi, Eur. Phys. J. C \textbf{81}, 490 (2021). %\textit{Imprints of dark matter on black hole shadows using spherical accretions}

\bibitem{Kiselev:2003} 
V.~V.~Kiselev, 
Class.\ Quantum Grav.\ {\bf 20}, 1187 (2003). %Quintessence and black holes


\bibitem{Kiselev:2003dm} 
V.~V.~Kiselev, 
arXiv:gr-qc/0303031. %Quintessential solution of dark matter rotation curves and its simulation by extra dimensions

\bibitem{Li:2012} 
M.-H.~Li and K.-C.~Yang, 
Phys.\ Rev.\ D {\bf 86}, 123015 (2012). %Galactic dark matter in the phantom field

\bibitem{Jha2025}
S. K. Jha, JCAP \textbf{09}, 069 (2025). % \textit{Black hole surrounded by perfect fluid dark matter with a background Kalb-Ramond field}


\bibitem{Pani2011} 
P.~Pani, V.~Cardoso, and T.~Delsate, 
Phys.\ Rev.\ Lett.\ {\bf 107}, 031101 (2011).



\end{thebibliography}
\end{document}